\documentclass[Letter]{aa}

\usepackage{graphicx}
\usepackage{physics,amsmath}
\usepackage{txfonts}
\usepackage{xcolor}
\usepackage{caption}
\usepackage{subcaption}
\usepackage{float}

\usepackage{natbib}
\usepackage{lastpage}
\bibpunct{(}{)}{;}{a}{}{,}
\usepackage[colorlinks=true, allcolors = blue]{hyperref}
\usepackage{ulem}

\begin{document}

    \title{Effects on $\varv \sin i$ determinations of O stars from 3D model atmospheres with high turbulent velocities}

    \author{L. Delbroek
          \and
          J.O. Sundqvist
          \and
          F. Backs
          \and
          T. Ceulemans
          \and
          D. Debnath
          \and
          P. Schillemans
          }

   \institute{Institute of Astronomy, KU Leuven, Celestijnenlaan 200D, bus 2401, 3001 Leuven, Belgium\\
             }

   \date{Received 27 March 2026/ Accepted 27 April 2026}
 
  \abstract  
   {When studying massive stars and their life cycles, rotation plays a key role. Hence, understanding the rotation of these stars is crucial when determining their properties, or for constraining evolutionary models.} 
   {We examined the reliability of the standard methods to derive projected rotation speeds $\varv \sin i$ from photospheric spectra of hot massive stars in the presence of high turbulent velocities.}
   {We included rotation in the spectral synthesis of O stars by means of three-dimensional (3D) model atmospheres showing significant photospheric turbulent velocities. We then used these as mock observations to back-test the Fourier transform (FT) and goodness-of-fit (GOF) methods commonly used for empirical determination of $\varv \sin i$ when the turbulent velocity field is not known.} 
   {When the expected $\varv \sin i > \varv_{\rm macro}$, with $\varv_{\rm macro}$ the macroturbulent velocity, FT $\varv \sin i$ determinations (most of the time) give reasonable results. However, if $\varv \sin i < \varv_{\rm macro}$ the method is no longer reliable. Results from the GOF method show that if one parameter is not significantly larger than the other, empirical best-fit values may be located in a large region of the $\varv \sin i$ -- $\varv_{\rm macro}$ parameter space, independent of the true values. The degenerate region closely follows the shape of the empirical formula found by \citet{Howarth07}, $C = \sqrt{\varv_{\rm macro}^2/2+(\varv \sin i)^2/4}$, with $C_i \ge C_e$ for $C_i$ the inferred value and $C_e$ the expected.}
   {Our analysis shows clearly that, generally, only the sum $\sqrt{\varv_{\rm macro}^2/2+(\varv \sin i)^2/4}$ can be approximated through the standard spectroscopic methods used to infer these parameters individually. Only in the case where one of the two clearly dominates, can good constraints on the dominating parameter be derived. This demonstrates that previously found empirical correlations between $\varv \sin i$ and $\varv_{\rm macro}$ as well as the derived statistical distributions of the observed rotation rates for O-star populations will need to be re-analysed and re-interpreted.}

   \keywords{Stars: massive -- Stars: atmospheres -- Stars: winds, outflows -- Methods: numerical -- Hydrodynamics -- Radiative transfer}

   \maketitle

\section{Introduction}

 The influence of rotation on massive stars already begins during the initial collapse of a rotating molecular cloud \citep{Bodenheimer_1995, rot_2011, rot_2016}, and lasts up to their brutal deaths as a core-collapse supernova or hypernova (\citealt{Supernovae_rot_2014}). During its life as a main-sequence star, and later on during the evolved stages, the effects of rotation cannot be ignored (\citealt{Maeder_2000, Maeder_2012}).

Even though rotation influences the entire life cycle of a massive star, we cannot directly measure the rotation rate in most cases; an exception to this is a small number of strongly magnetic massive O stars (e.g. \citealt{Sundqvist13}). Hence, the projected stellar rotation, $\varv \sin i$, is commonly deduced ($i$ being the inclination angle, in this work sometimes also referred to as $\alpha$) from the observed broadened line spectrum. This approach, however, has one primary issue, which is the potential degeneracy between the rotational line broadening and the other broadening processes in the line-formation regions. From observations of absorption lines in O stars, it has long been known that a wide range of velocities must be present in their atmospheres \citep{Slettebak56, macroturb_Conti}. The lines are much broader (typical velocity dispersions $\sigma_{\rm V} \sim 50$ km/s) than implied by thermal motions, and the observed line shapes cannot be explained by stellar rotation alone \citep{macroturb_Conti,Howarth_1997,Ryans_2002,Simon-Diaz_2007,Lefever_2007,Sim_n_D_az_2010, Nadya2024}. To overcome the issue of separating rotational broadening from this `turbulent' contributor, empirical techniques such as the Fourier transform (FT) method, which looks for the first zero in the Fourier power spectrum to determine $\varv \sin i$ (see Sect. \ref{FT method}), and the goodness-of-fit (GOF) method (see Sect. \ref{GOF method}), which uses the different shapes of the rotational and turbulent broadening to find $\varv \sin i$ through a fitting algorithm, have been utilised \citep{Gray_2005, Simon-Diaz_2007}. However, \citet{Sundqvist_2014} showed that these methods may become unreliable at $\varv \sin i \la 50$~km/s due to the degeneracy of the broadening components (see also \citealt{Howarth07, Markova14}). 

Recent spectral synthesis directly from 3D model atmospheres \citep{Schultz_2023, First_paper_Lara} shows that this `extra broadening' indeed is naturally reproduced by the high turbulent velocities present in such simulations of O-star photospheres.\footnote{This mimics results previously found for the Sun \citep{Asplund_2009}, however the characteristic turbulent velocities are more than an order of magnitude larger for O stars.} These models, however, did not include rotation and thus the question regarding separation between turbulent and rotational broadening could not be addressed. In this letter we introduce rotation in the spectral synthesis of our 3D models and use this to analyse the total broadening as well as backtrack the reliability of the empirical standard 1D methods.

\section{Methods} \label{Methods}
The fundamental parameters for the O-star model studied in this work can be found in Table \ref{table:Models}. For the details and methodology of the synthetic line profile calculations, more specifically, the approximate non-local thermodynamic equilibrium\footnote{This aNLTE method \citep{Puls_2000} corrects level population numbers computed in local thermodynamic equilibrium, by using (potentially) different radiation and gas temperatures, as well as a modified spherical dilution factor. This dilution factor is an approximate way of accounting for sphericity when computing occupation numbers and source functions, which is reasonable for modelling the optical absorption lines considered in this work.} and 3D radiative transfer techniques (synthetic spectra computed from a formal solution using long characteristics in a cylindrical coordinate system, pointed at the observer) and the unified model atmosphere with wind models, we refer to \citealt{First_paper_Lara}.  We computed the line profiles directly from the 3D simulations, focusing on the typical optical absorption lines used to infer $\varv \sin i$ and macroturbulence in O stars. The 3D RHD (radiation hydrodynamics) model has been mapped out to cover the complete atmosphere and wind volume using the same technique as in \citet{First_paper_Lara}\footnote{We called this a mixed-sphere model, as it is a spherical model that uses many different snapshots-boxes during its construction. See \citet{First_paper_Lara} for more details.}. We added rotation to our 3D models utilising azimuthal angular momentum conservation outside a fixed inner boundary radius with rotation $\varv_{\rm rot,in}$ (see Appendix \ref{Adding_rotation_method}, here referred to as the inner boundary rotation method). We note that this approach neglects dynamical feedback effects from the rotation upon the atmospheric density and temperature structure, which should be a reasonable approximation as long as the star is not rotating too rapidly \citep[e.g.,][]{Hillier12, Sundqvist12}. We used radial averages in the static limit to mimic what would come out of an equivalent 1D model photosphere analysis; to reproduce the equivalent widths of the 3D-based lines, such 1D-based lines must also introduce microturbulence. For the O\,III 5594 $\AA$ in focus here, one needs $\varv_{\rm mic} = 15$ km/s, (\citealt{First_paper_Lara}). With these new sets of line profiles, we then back-tested to see if we could find the correct $\varv \sin i$ values using the FT and GOF methods. More specifically, we added rotation to one 3D model and calculated the O\,III 5594 $\AA$ line flux profile for different inclination angles. We then used these profiles as mock observations to derive $\varv \sin i$ and macroturbulence using the standard GOF and FT methods. Two remarks that should be made are the following: i) In the 3D synthetic line profiles we do not have the same noise levels as one would expect from real observations; ii) The 3D line profiles are not necessarily symmetric, and can have irregularities. This cannot be reproduced by the 1D models and methods, which may impact the results somewhat.

\begin{figure} [ht]
    \centering
    \includegraphics[width=0.9\linewidth]{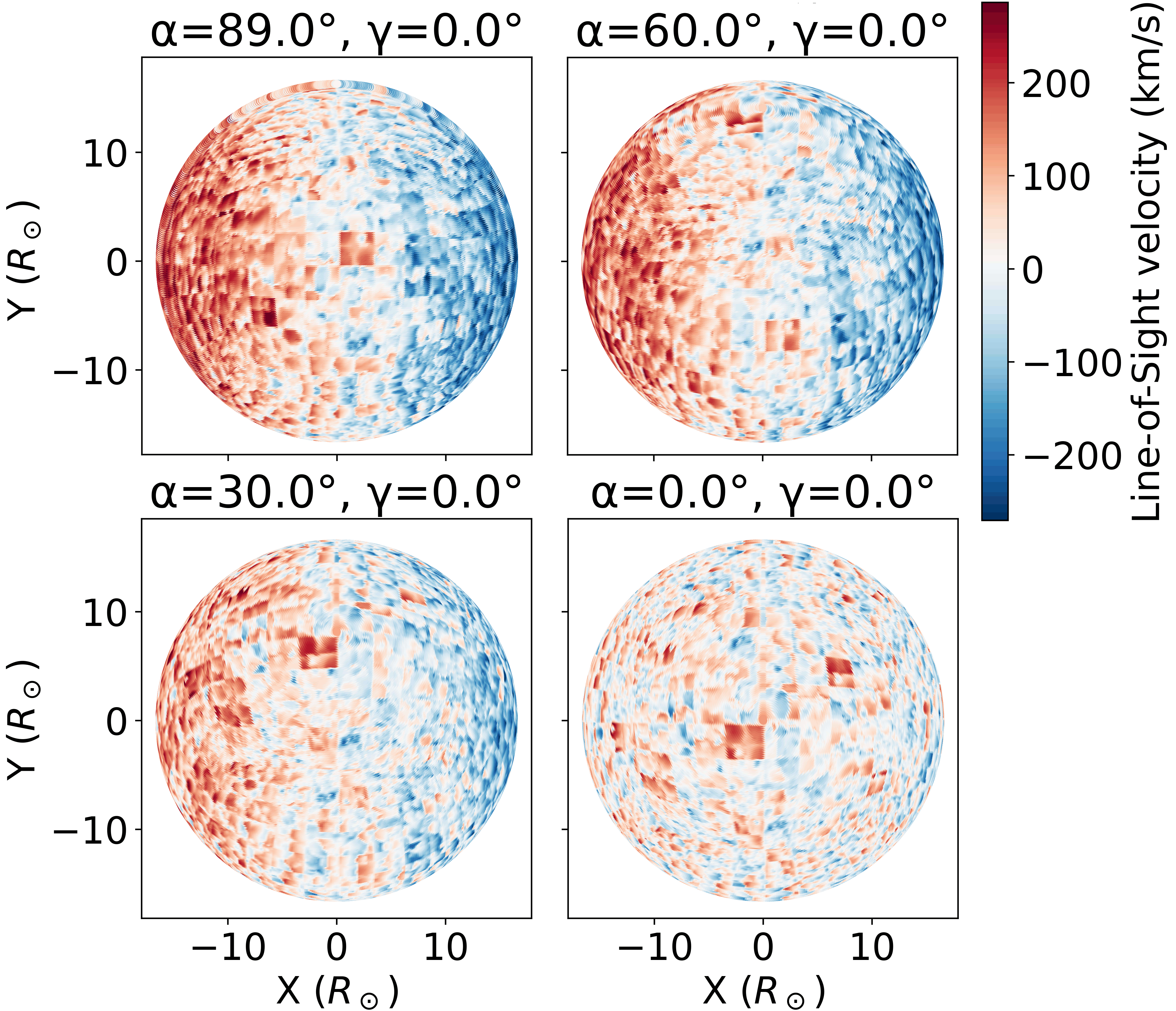}
    \caption{Visualisation of LoS velocities (at the optical photosphere) for different observer positions. The rotation at the inner boundary, for the model shown here, is $\varv_{\rm rot,in} = 200$~km/s, which corresponds to a rotation at the photosphere of approximately 190~km/s. The $\alpha$ angle is the inclination; the $\gamma$ angle the rotation around the equator.}
    \label{fig:rot_visualisation}
\end{figure}

\begin{figure} [ht]
    \centering
    \includegraphics[width=0.9\linewidth]{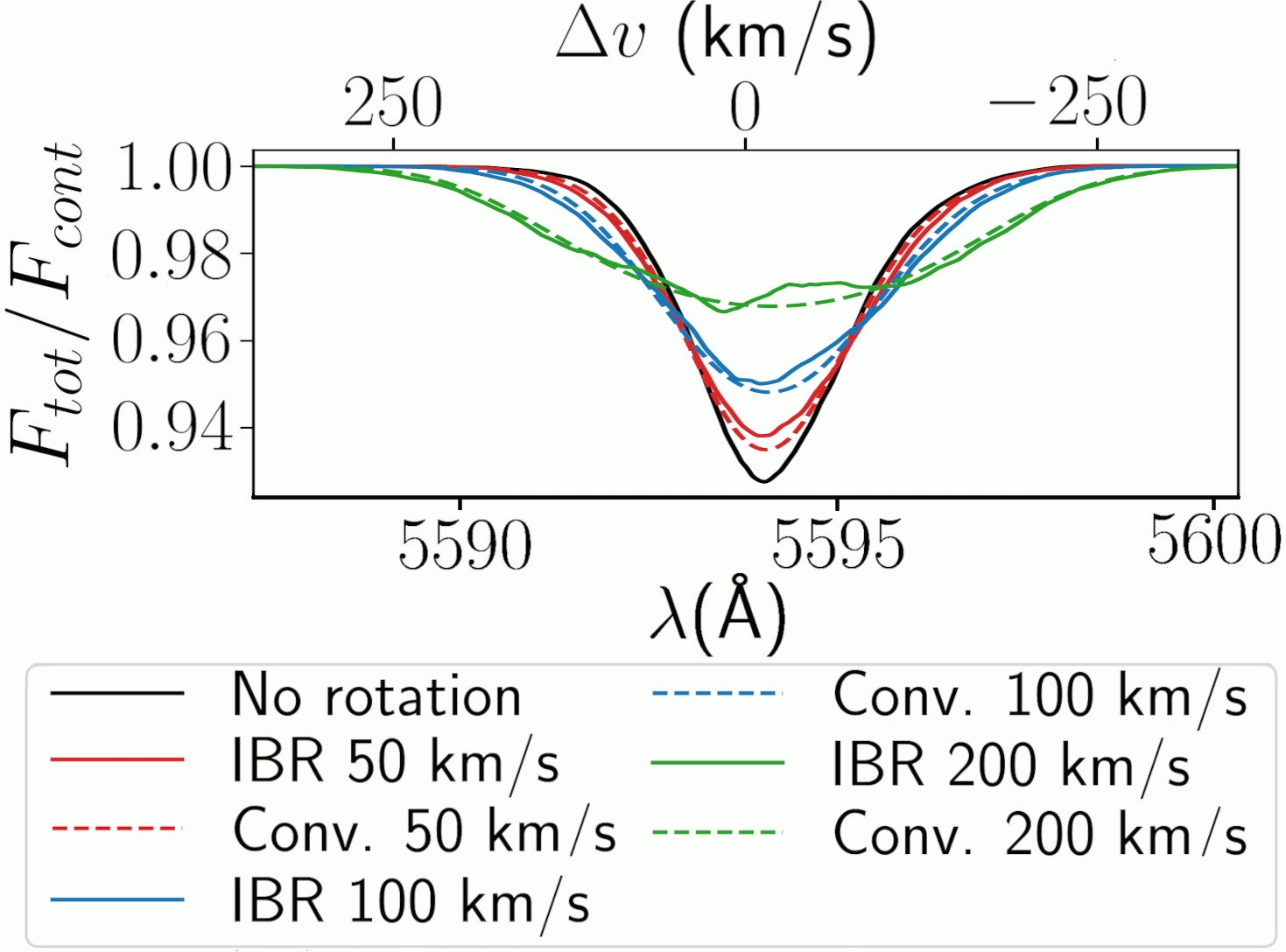}
    \caption{O III 5594 $\AA$ line from a 3D O-star model atmosphere, calculated using different methods for adding rotation. The black line corresponds to the non-rotating model, while the coloured lines correspond to lines being calculated with rotation using two different methods (IBR = inner boundary rotation, Conv. = convolution).}
    \label{fig:rot_conv_plot}
\end{figure}

\section{Results}

Adding rotation to the 3D model gives line-of-sight (LoS) velocities as shown in Fig.~\ref{fig:rot_visualisation}. In the top left panel, at an inclination $i = \alpha=89^\circ$, a LoS velocity gradient is visible from left to right, illustrating the stellar rotation. Additionally, the turbulent motions in the model are visible. For lower values of $\alpha$ the rotation becomes less visible and disappears completely for the pole-on view $\alpha=0^\circ$. Although the systematic gradient disappears, a large spread in velocities remains due to the atmospheric turbulence. We have some snapshots in our spherical model that show high LoS velocities across nearly the entire snapshot surface. As we are interested here in surface-averaged properties, however, these individual snapshot variations level out. In Fig. \ref{fig:LoS_appendix}, an average LoS velocity at the optical photosphere for individual snapshots used to construct our spherical models is shown. Three synthetic line profiles including rotation at different rates (for an equatorial observer) are shown in Fig. \ref{fig:rot_conv_plot}. This figure also shows line profiles first calculated for the turbulent non-rotating model, and then simply convolved with a $\varv \sin i$ following \citep{Gray_2005}. The figure shows that this convolution method actually gives good results for photospheric absorption lines, as long as the turbulent velocity field is known beforehand (which it is here from the 3D model itself). Below we now test if this holds, also when the turbulent velocities are not known, but are instead treated with an additional fit parameter or ignored (in the case of the FT method).  

The key results are summarised in Fig. \ref{fig:all_results_combined}, where, for each method, we show the determined values for $\varv \sin i$ (using both GOF and FT) and $\varv_{\rm macro}$ (only GOF), and compare them to the expected values from the 3D-based mock observation. The expected value for isotropic Gaussian macroturbulence is $\varv_{\rm macro} = 74$~km/s, as derived from the non-rotating 3D model (see also \citealt{First_paper_Lara}). If $\varv \sin i$ is significantly higher than $\varv_{\rm macro}$, we get decent $\varv \sin i$ determinations, as seen on the left-hand sides of the top and bottom panels. The opposite is true if $\varv \sin i \approx \varv_{\rm macro}$. Finally if $\varv \sin i \ll \varv_{\rm macro}$, FT and GOF determinations of $\varv \sin i$ become much too high, showing overestimates up to a factor of 3 for the GOF and even more than a factor of 10 for the FT method (right-hand sides of the top and bottom panels). On the other hand, in this latter regime $\varv_{\rm macro}$ is well constrained. By contrast, if $\varv _{\rm macro} < \varv \sin i$, $\varv_{\rm macro}$ determinations are generally poor. This shows that we can only obtain good constraints on the projected stellar rotation if $\varv \sin i$ is (much) larger than $\varv_{\rm macro}$, and vice versa for macroturbulence. 

\subsection{Fourier transform $\varv \sin i$ determination}\label{FT method}

\begin{figure} [h]
    \centering
    \includegraphics[width=0.9\linewidth]{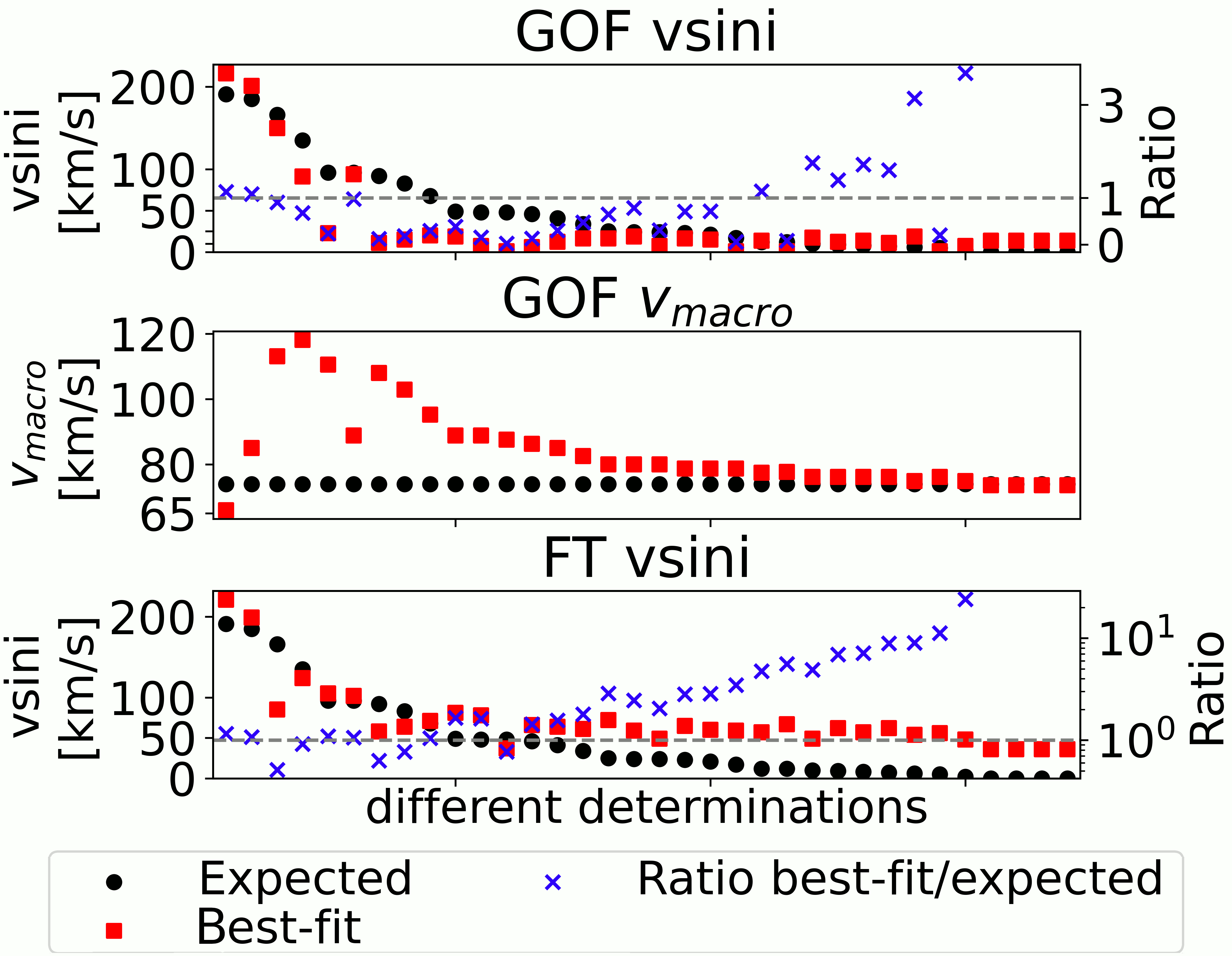}
    \caption{Results for $\varv \sin i$ and $\varv_{\rm macro}$ GOF and FT determinations (Best-fit) and the corresponding expected values (Expected). For the $\varv \sin i$ results, we also show the ratio of determined to expected values. The grey dotted line indicates where the ratio is equal to one. }
    \label{fig:all_results_combined}
\end{figure}

\begin{figure} [h]
       \centering
\includegraphics[width=0.9\linewidth]{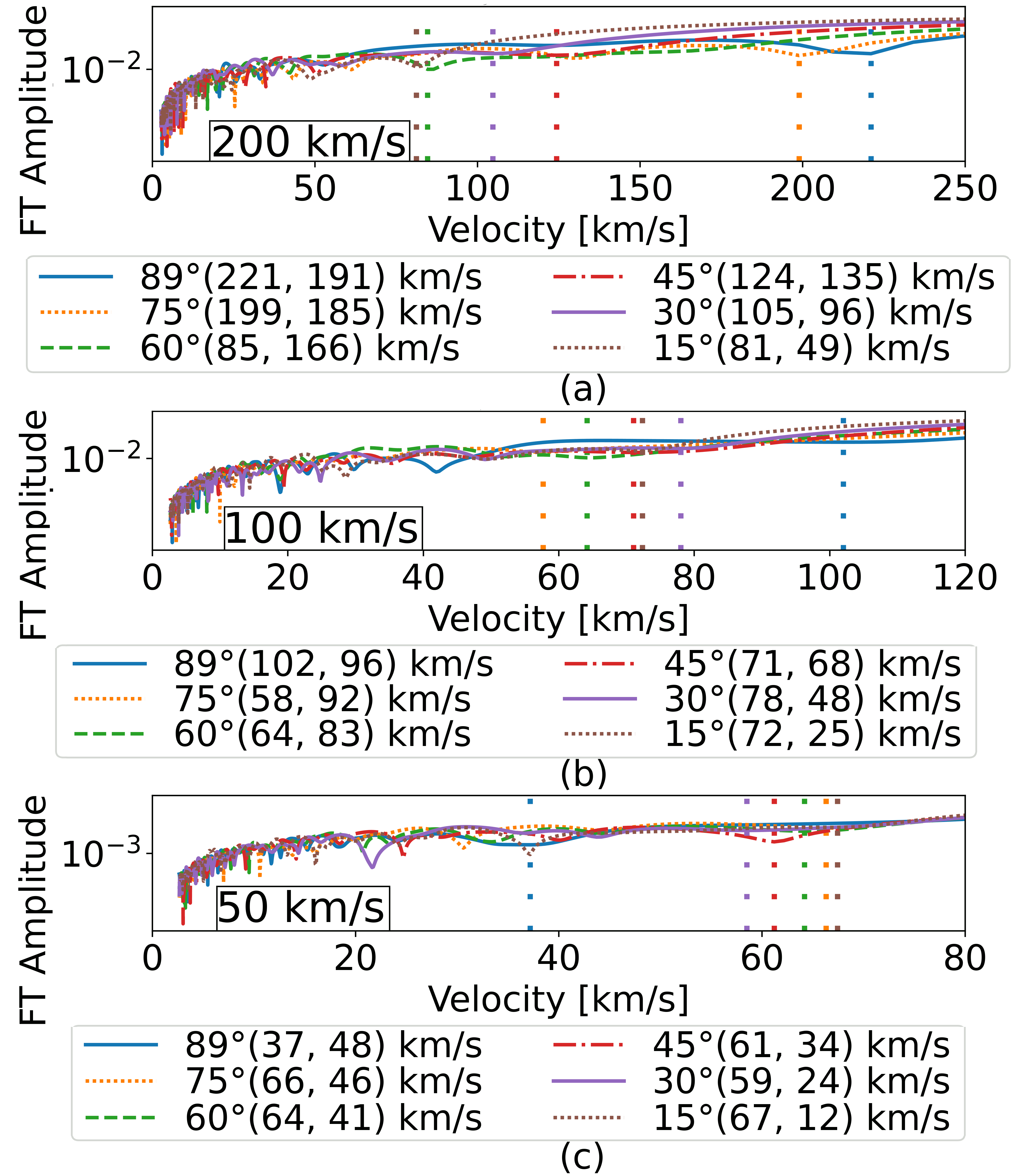}
     \caption{$\varv \sin i$ determined by the FT method for different $\varv_{\rm rot,in}$ (indicated in the lower left corner of each panel); angles are inclination angles $i$ of the observer. The first number in the brackets displays $\varv \sin i$ found by the FT method, also indicated in the figure by the vertical dotted lines, whereas the second number is the expected value at the photosphere.}
         \label{FT_4_panel} 
   \end{figure}

The FT $\varv \sin i$ results are shown in Fig. \ref{FT_4_panel}. The determination was also done for the pole-on view (not included in the figure), for which all determinations gave $\varv \sin i \ga $30~km/s. The first thing to note is that the FT method always gives a non-zero result, indicating that fiducial minima are created by the presence of the high turbulent velocities. Figure \ref{FT_4_panel}a and \ref{FT_4_panel}b show that as long as $\varv \sin i \ga \varv_{\rm macro}$, the FT method produces reasonable results. However, even in this range there is significant scatter, such that some best fits are less than satisfactory (e.g. top panel with $i = 60^\circ$). If $\varv \sin i < \varv_{\rm macro}$, the measurements become much less reliable. This is seen directly in Fig. \ref{FT_4_panel}c, where all measurements except at the equator overestimate $\varv \sin i$ by at least 40 percent and up to a factor of five, depending on which inclination angle is considered. This shows that one must be very careful when using this method, especially when the macroturbulent velocity is not known in advance, as also visible in Fig. \ref{fig:all_results_combined} (in particular, the right-hand side of bottom panel).

    \subsection{Goodness-of-fit determination} \label{GOF method}

\begin{figure} [ht]
       \centering
\includegraphics[width=0.9\linewidth]{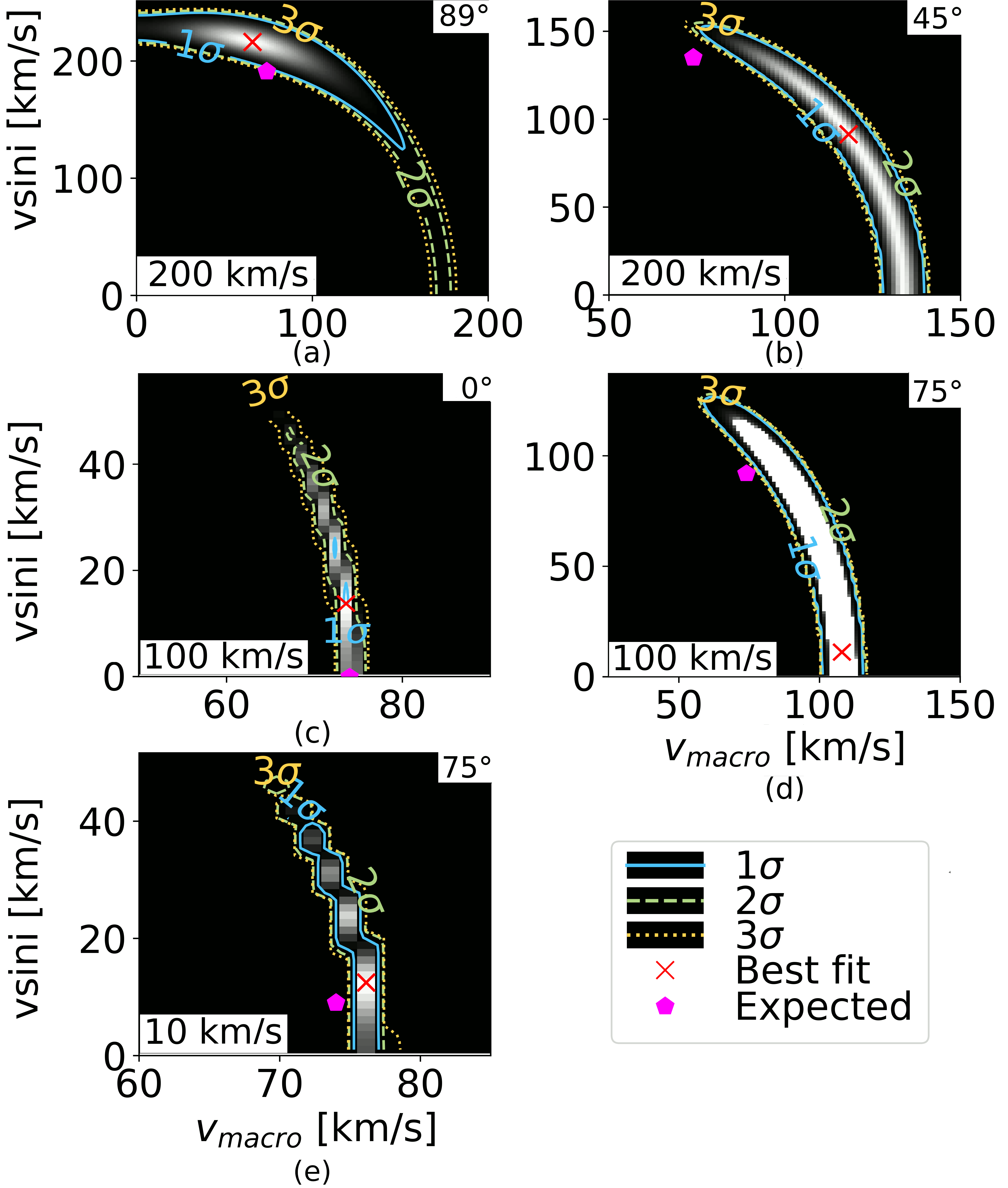}
      \caption{$\varv \sin i$ and $\varv_{\rm macro}$ determined by the GOF method. The observer inclination angle $i$ is shown in the top right corner of the panels, and $\varv_{\rm rot,in}$ in the lower left corner. The red cross indicates best-fit values, while the magenta pentagon shows what the expected values are. We note that the colour scaling of this chi-squared distribution function is in log space, not linear space.}
         \label{GOF_panel} 
   \end{figure}

For our macroturbulent velocity determinations we assume an isotropic Gaussian, not the radial-tangential method \citep{Gray_1975}. This gives a straightforward connection to the velocity dispersion $\sigma_{\rm V} = \varv_{\rm macro}/\sqrt{2}$ measured in the RHD models \citep{First_paper_Lara}. To compare to observational results, we further need to estimate the variance of our synthetic data. For this we assume errors are independent and identically distributed, following a normal distribution with the region defined by the full width at half maximum (FWHM) of the line. The total standard deviation is then estimated by the standard deviation of the residual within the FWHM region, and is used to determine our chi-squared distribution function and derive confidence intervals. This method is far from perfect, and the confidence intervals should be viewed as approximations only. Even so, the best-fit values should remain unaffected and we focus on the general trends of our results. In Fig. \ref{GOF_panel} every panel has a curved white shape in which the best-fit value lies. Depending on the rotation rate and observation angle, this shape (within our confidence intervals) is more focused or stretches over a larger range. When placing the observer at the equator for $\varv_{\rm rot,in} = 200$ km/s (Fig. \ref{GOF_panel}a), the best-fit region is small (with the $\sigma$ contours still outlining the full region), while $i = 75^{\circ}$ and $\varv_{\rm rot,in} = 100$ km/s (Fig. \ref{GOF_panel}d) has a much larger banana-shaped best-fit region. Moreover, even for a polar observer we derive $\varv \sin i > 0$; from Fig. \ref{GOF_panel}c we see that the $1\sigma$ regions do not actually extend to $\varv \sin i=0$.

\citet{Howarth07} empirically found that the broadening of an observed spectral line may be described as
\vspace{-5pt}
\begin{equation}
\sqrt{\sigma_V^2 + \left(\frac{\varv \sin i}{2}\right)^2} = C,
\label{Eq:gof_total}
\end{equation}

with $C$ being a constant. Comparing our GOF results to this formula (illustrated in Fig. \ref{fig:GOF_plot_appendix}) shows that when using our best-fit result to construct a `Howarth curve', this curve very nicely matches our found shapes for the best-fit values; that is, for the GOF method, the best-fit value lies in a region covered by Eq. \ref{Eq:gof_total} and it can lie anywhere in this region. The smaller this region is, the higher the chance that we get an accurate $\varv \sin i$ determination. Hence, the GOF method can give reliable results when $\varv \sin i \gg \varv_{\rm macro}$. If this is not true, one should be very careful in trusting GOF results, especially when one does not have any other constraints on $\varv_{\rm macro}$. In Figures \ref{GOF_panel} and \ref{fig:GOF_plot_appendix} it can be seen that the expected values do not always lie in the best-fit $3\sigma$ region. Furthermore, the best-fit Howarth curve determines an upper limit for the value of C, since the expected Howarth curve is located at or below the best-fit Howarth curve ($C_i \ge C_e$ for $C_i$ the inferred value and $C_e$ for the expected value). This shows that errors may be larger than estimated from confidence intervals, which can cause extra complications in the $\varv_{\rm macro}$ and $\varv \sin i$ determinations. It should be noted here that the $\varv_{\rm macro}$ determinations are good if $\varv_{\rm macro}$ dominates ($\varv \sin i \ll \varv_{\rm macro}$).

\section{Conclusions} \label{Disc_and_conc}
Both the FT and GOF methods have severe limitations for determining $\varv \sin i$ in the presence of high atmospheric turbulent velocities. If $\varv \sin i$ is not significantly higher than $\varv_{\rm macro}$, an accurate determination of the projected rotation rate is only possible if $\varv_{\rm macro}$ can be calibrated independently with sufficient confidence (for example from 3D atmosphere simulations). If not known independently, the macroturbulent velocity can only be empirically determined if it dominates the broadening. This may significantly alter previous determinations of observed rotation rates for O stars, and a range of previous results must likely be re-analysed and re-interpreted. This includes fiducial correlations between $\varv \sin i$ and $\varv_{\rm macro}$ \citep{Simon-Diaz2017} as well as the statistical distribution of observed rotation speeds in O-star populations \citep{Ramirez13}. 

\begin{acknowledgements}

The computational resources used for this work were provided by Vlaams Supercomputer Centrum (VSC) funded by the Research Foundation-Flanders (FWO) and the Flemish Government. The authors gratefully acknowledge support from the European Research Council (ERC) Horizon Europe under grant agreement number 101044048, from the Belgian Research Foundation Flanders (FWO) Odysseus program under grant number G0H9218N, from FWO grant G077822N, and from KU Leuven C1 grant BRAVE C16/23/009. The authors would like to thank the previous and current members of the KUL-EQUATION group for their contributions, fruitful discussions, and suggestions. We made significant use of the following packages to analyse our data: {\fontfamily{qcr}\selectfont NumPy} \citep{harris_2020}, {\fontfamily{qcr}\selectfont SciPy} \citep{virtanen_2020}, {\fontfamily{qcr}\selectfont matplotlib} \citep{hunter_2007}, {\fontfamily{qcr}\selectfont Python amrvac\_reader} \citep{keppens_2020}.

\end{acknowledgements}

\bibliographystyle{aa}
\bibliography{references}

\begin{appendix}
    
\setcounter{table}{0}
\renewcommand{\thetable}{A\arabic{table}}
\setcounter{figure}{0}
\renewcommand{\thefigure}{A\arabic{figure}}
\section{Adding rotation to models}  \label{Adding_rotation_method}
    In order to verify how well the FT and GOF method are able to find an accurate rotation rate, we first need to include the rotation in our models. This is done in the following way, using conservation of angular momentum (see e.g. \citealt{Muller_2014}) for each point in our grid, which is in the spherical coordinate system $(r,\theta, \phi)$ with r in range $(R_0, R_{max})$, $\theta$ in range $(0, \pi)$ and $\phi$ in range $(0, 2\pi)$:

    \begin{align}
        \varv_{\rm rot} &=  \varv_{\rm rot, in}\sin(\theta) \bigg( \frac{R_0}{r} \bigg) \\
        \varv_x &= \varv_x - \varv_{\rm rot}\sin(\phi) \\
        \varv_y &= \varv_y + \varv_{\rm rot}\cos(\phi) \\
        \varv_z &= \varv_z
    \end{align}
      Here, $ \varv_{\rm rot}$ is the calculated rotation at a specific grid point, $\varv_{\rm rot, in}$ is the rotation at the inner boundary (on the equator), $\varv_x$ is the x-component of the velocity field, $\varv_y$ is the y-component of the velocity field and $\varv_z$ is the is the z-component of the velocity field. Since we work with $\varv_{\rm rot, in}$, the rotation at the inner boundary, we also refer to this method as the `inner boundary rotation' method. In Fig. \ref{fig:rot_conv_plot}, we compare this method for adding rotation to convolution, using the methods described in \citet{Gray_2005}.

\section{Additional Table}
\renewcommand{\thetable}{B\arabic{table}}
\begin{table} [h]
\caption{Fundamental parameters for the $\langle 3\rm{D} \rangle$ O-star model studied in this work.}       
\label{table:Models}      
\centering          
\begin{tabular}{c | c c c}  
\hline\hline       
Model & $\left<T_{\rm eff} \right> \rm [kK]$ & $M_\star/M_\odot$ & $\langle R_\star \rangle/R_\odot$ \\
\hline                    
$\rm{O}4$ & 40.2  & 58.3 & 16.2 \\
\noalign{\vskip 0mm}\cline{2-4}\noalign{\vskip 0.75mm}\cline{2-4}

& $\log_{10} \left(\left<L_\star\right>/L_\odot\right)$ & $\left<L_\star\right>/L_{\rm edd}$ & $\log_{10} \left<g_\star\right>$ \\
\cline{2-4}

& 5.79 & 0.27 
& 3.79 \\
\noalign{\vskip 0mm}\cline{2-4}\noalign{\vskip 0.75mm}\cline{2-4}

& $\log_{10} \left<\Dot{M}\right>\ [M_\odot/yr]$ & $\left< \varv_{\rm max} \right> \rm [km/s]$ & \\
\cline{2-4}

& -5.90 & 1800 & \\
\hline                  
\end{tabular}
\tablefoot{From left to right, columns display model name, effective temperature, stellar mass, radius, luminosity, Eddington ratio, surface gravity, mass-loss rate, and maximum velocity. Angle brackets denote averaged quantities as explained in text. Table adapted from \citealt{First_paper_Lara}.}
\end{table}

\newpage
\section{Figures}  
\renewcommand{\thetable}{C\arabic{table}}
\renewcommand{\thefigure}{C\arabic{figure}}

\begin{figure} [!h]
     \centering   
     \includegraphics[width=0.97\linewidth]{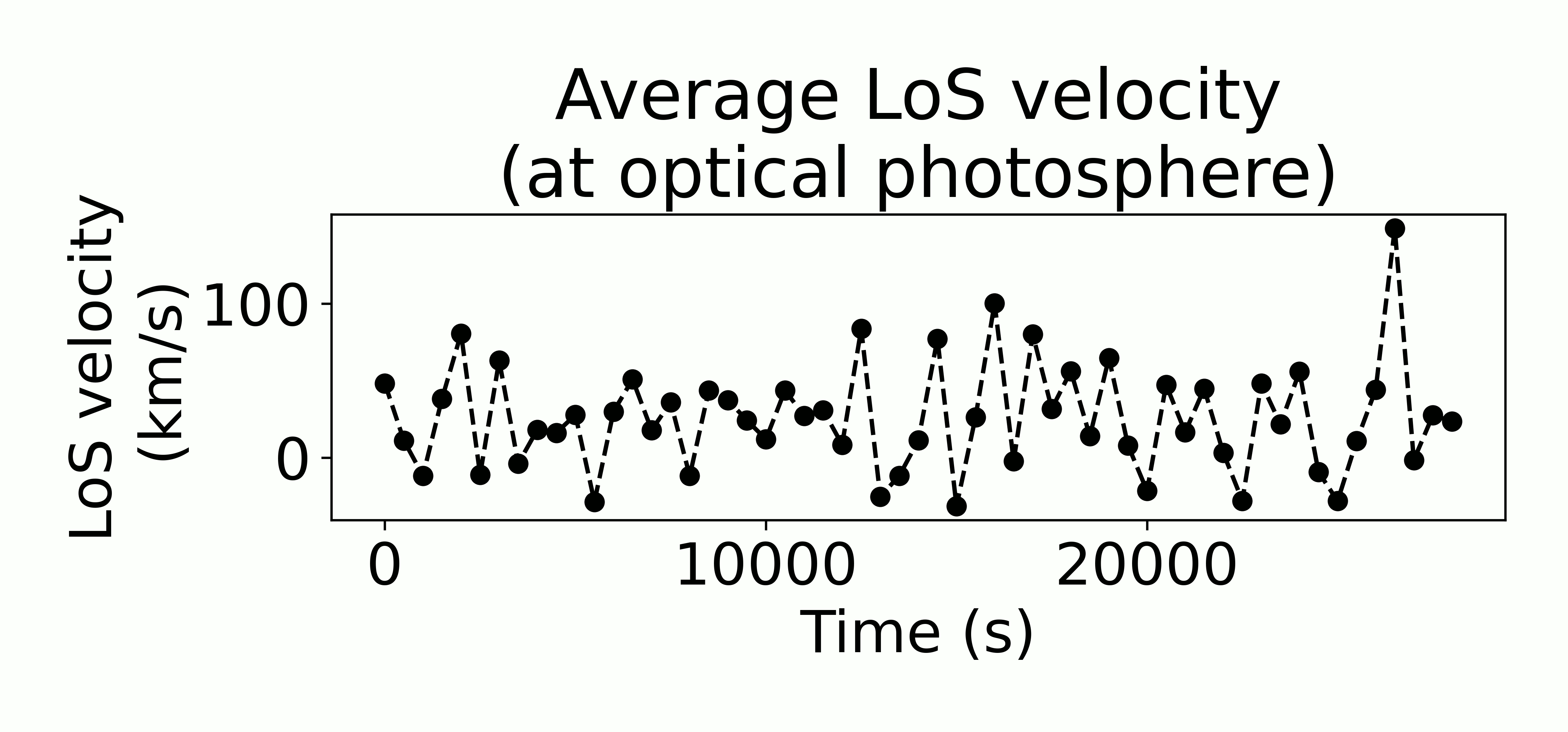}
    
     \caption{Average LoS velocities at the optical photosphere of individual snapshots, used to construct our spherical models. Each snapshot is separated by approximately 500 seconds in time, showing the variability of these simulations in time. The positive LoS velocities correspond to a general inward-falling motion at the photosphere, while a negative LoS velocity corresponds to an outward (expanding) motion. We note that these are averages, and this does not mean that all matter is moving inward or outward. In Fig. \ref{fig:rot_visualisation}, it might sometimes appear that way, due to the lateral resolution of the snapshots being reduced in order to limit the number of grid points and keep the calculation computationally possible (we already have $N_{tot} \sim 10^8$ in our full 3D volume). Hence, an (on average) infalling or expanding motion might appear omnipresent at this specific optical depth, even though this is not the case in reality. As can be seen in \citet{First_paper_Lara}, the variability of these snapshots levels out when doing line synthesis for many time steps. This becomes clear when comparing the spectral synthesis of 'mixed-sphere' models, to the average of all spectral lines constructed from mapping out one snapshot on the full sphere. Another important point to note is that our spectral lines are not formed at exactly the average photosphere, but rather at multiple layers and/or radial distances. In consequence, one cannot simply look at the LoS velocity at the optical photosphere, and form direct expectations about the formed spectral lines from them.}
     \label{fig:LoS_appendix}
 \end{figure}

\begin{figure} [!h]
     \centering   
     \includegraphics[width=0.97\linewidth]{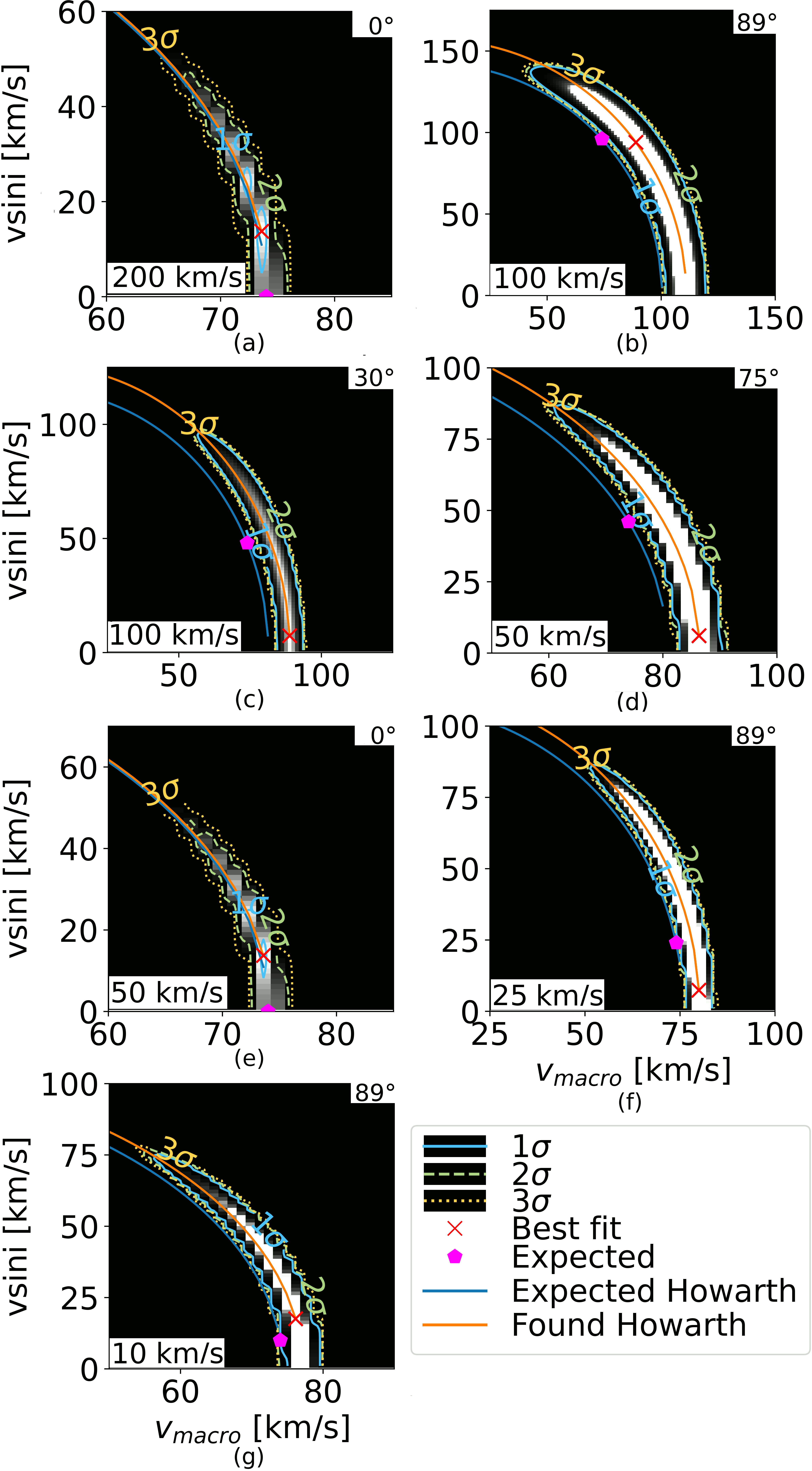}
    
     \caption{$\varv \sin i$ and $\varv_{\rm macro}$ determined by the GOF method, applied on the O III 5594 $\AA$ line results from a 3D RHD model that has been mapped out to cover the complete atmosphere and wind volume using the same technique as in \citet{First_paper_Lara}. The observer inclination angle $i$ is shown in the top right corner of the panels and $\varv_{\rm rot,in}$ in the lower left corner. The red cross gives the best-fit values for $\varv \sin i$ and $\varv_{\rm macro}$, based on the GOF method, while the magenta pentagon shows what the expected values are. The blue and red curves correspond to the empirically found formula for broadening of an observation by \citet{Howarth07}, which is applied here on our results. The blue curve corresponds to using our expected results, while the red corresponds to using our found best-fit value results. We note that the colour scaling of this chi-squared distribution function is in log space, not linear space.}
     \label{fig:GOF_plot_appendix}
 \end{figure}

\end{appendix}

\end{document}